\documentclass[conference]{IEEEtran}

\usepackage[english]{babel}
\usepackage{courier}
\usepackage{amstext}
\usepackage{times}
\usepackage{helvet}
\usepackage{courier}
\usepackage[cmex10]{amsmath}
\usepackage{amssymb}
\usepackage{latexsym}
\usepackage{eurosym}
\usepackage{wasysym}

\makeatletter
\newif\if@restonecol
\makeatother

\usepackage{etex}
\usepackage{tikz}
\usepackage{pslatex}
\usepackage[arrow, matrix, curve]{xy}
\usepackage{epsfig}
\usepackage{pstricks,egameps}

\begin{document}

\title{Bewelcome.org -- a non-profit democratic hospex service set up for growth.}

\author{\IEEEauthorblockN{Rustam Tagiew}
\IEEEauthorblockA{Alumni of\\TU Bergakademie Freiberg\\
yepkio@mail.ru}}

\maketitle

\begin{abstract}
This paper presents an extensive data-based analysis of the non-profit democratic hospitality exchange service bewelcome.org. We hereby pursuit the goal of determining the factors influencing its growth. It also provides general insights on internet-based hospitality exchange services. The other investigated services are hospitalityclub.org and couchsurfing.org. Communities using the three services are interconnected -- comparing their data provides additional information.
\end{abstract}

\IEEEpeerreviewmaketitle

\section{Introduction}

\indent The rite of gratuitous hospitality provided by local residents to strangers is common among almost all known human cultures, especially traditional ones. We can cite melmastia of Pashtun \cite[p. 14]{melmastia}, terranga of Wolof \cite[p. 17]{teranga} and hospitality of Eskimo \cite{eskimo} as examples of traditional hospitality. A stranger hereby receives shelter, food, protection and other help. In return, the guest is expected to contribute at least symbolically to his hosts well-being.\\
\indent In the modern world the share of tourists among the strangers grew. Further, we will term adventurous budget tourists as travelers and concentrate on them, since they are the only ones seeking for gratuitous hospitality. Because travelers can turn into local residents and vice versa, the idea of hospitality exchange abbreviated as \textit{hospex} became relevant. Hospex participants host and can have hosts. They are organized in a community, where gratuitous hospitality is not responded directly but by hospitality of another participant or rather member. Further, we will term everybody as member, who had a real-life interaction with another member.\\
\indent The first hospex service was initiated by Servas International starting in 1949 \cite{servas}. A hospex service basically administrates so-called profiles -- participant description, contact and location. It is disputed, whether different organizations create different communities or are just different services for one single international hospex community, since some people use more than one hospex service \cite{hcsite}. We don't discuss this topic.\\
\indent A hospex service is easier to maintain, once a central online database of profiles is set up. In addition to profiles, recording accommodation reports aka references or comments became popular for on-line hospex services. Such a database is accessible over a web- and/or app-based front-end. At first, such service called 'Hospex' was created in 1992 \cite{hospex}. The installations of other services followed. Hospitality Club abbreviated as HC became the biggest of them with over 100k profiles in 2006 \cite{hcpaper}. Later, HC got outdistanced by Couchsurfing abbreviated as CS. CS has now the biggest number of profiles -- over 3M \cite{cssite}. We have to underline here the difference between the terms \textit{user}, which is basically a profile owner, and \textit{member} -- not every website user had a real-life interaction with other members in order to be called a member. Members are a subset of users.\\ 
\indent Hospex services always started as non-profit, donation and volunteer-driven organizations, since making profit on gratuitous hospitality is considered unethical in modern culture. Nevertheless, business on gratuitous hospitality is common at least in mediation of au-pairs to host families \cite[eg.]{aupair}.\\ 
\indent Unfortunately, we can not access the data from CS -- the biggest hospex service. CS became a for-profit corporation in 2011 and shut down the access of public science to its data. Scientists possessing pre-incorporation CS data are prohibited to share it with third parties. HC never allowed access of public science to its data. Therefore, we can only rely on the data kindly mirrored for us by BW on 04.03.2014, the public Google search data and the mirrored statistics from the CS website in 2011. The original statistics of CS website are no longer available. For what it's worth, subsequent versions of the official CS statistics page presented figures heavily reduced and are clearly different from their statistics published in 2011.\\ 
\section{Terminology}
\begin{description}
 \item[\textbf{Traveler}]\hfill Adventurous budget tourist.\\
 \item[\textbf{Hospex}]\hfill Indirect exchange of gratuitous hospitality.\\
 \item[\textbf{Community}] \hfill People interconnected by fulfilled hospex.\\
 \item[\textbf{Member}] \hfill A person, who had real-life interaction with another member concerning the hospex community.\\
 \item[\textbf{Profile}] \hfill Description, contact and location of a person agreed to hospex.\\ 
 \item[\textbf{Service}] \hfill A communication hub, which facilitates hospex arrangements.\\ 
 \item[\textbf{User}]\hfill A person, who created a profile using a hospex service by signing up.\\
\end{description}
\section{Related Work}
\indent We can identify three categories of existing scientific publications claiming hospex as their subject matter -- non-data scientific articles, analysis of survey data, and analysis of CS data. Survey data gives insights into mindsets, but not into real behavior processes on hospex. Currently we know about 4 research teams, who used pre-incorporation CS data \cite{victor,danderkar,overgoor,lauterbach}. All papers written by these research teams solely concentrate on the aspect of trust among the CS users. Particularly, they don't provide general insights as aimed in this paper. Further, the correctness of their work can not be double-checked, because they are not allowed to share the data anymore.\\
\section{General development of hospex services}
\begin{figure}
\includegraphics[scale=0.5]{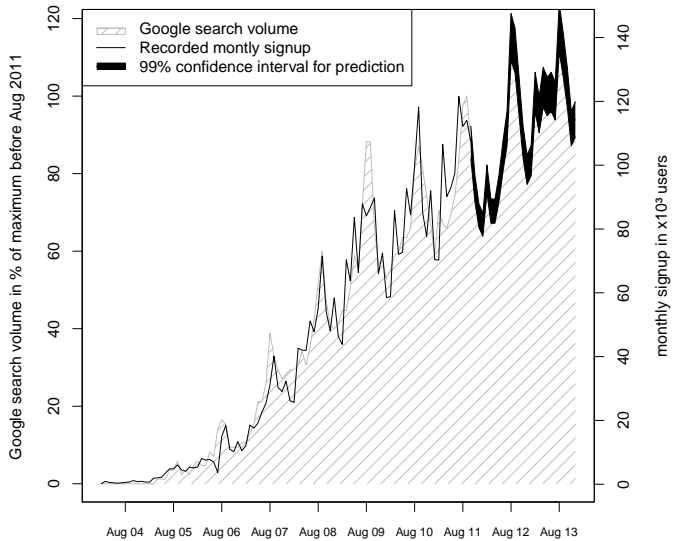}
\caption{High linear correlation of .971 between Google search volume for \glqq Couchsurfing\grqq and CS monthly signup in the years 2005--2011 allows linear prediction for the subsequent development.}
\label{secondf}
\end{figure}
\begin{figure}
\includegraphics[scale=0.5]{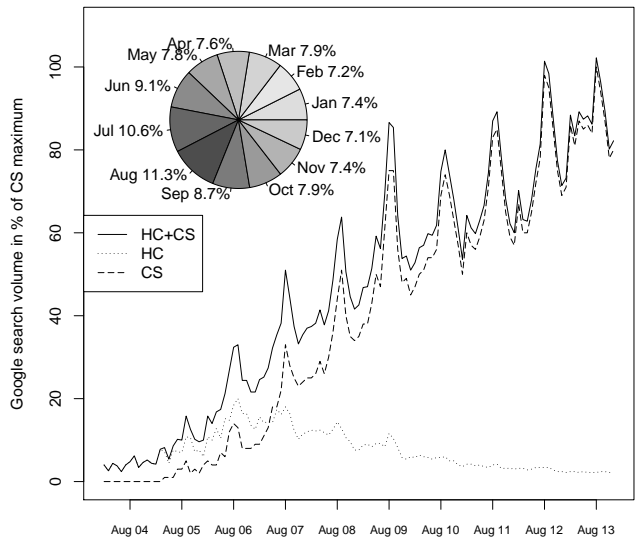}
\caption{Google search volume for HC, CS and HC+CS. Pie chart of growth adjusted mean distribution of Google search for HC+CS over months.}
\label{cskillhc}
\end{figure}
\begin{figure}
\includegraphics[scale=0.5]{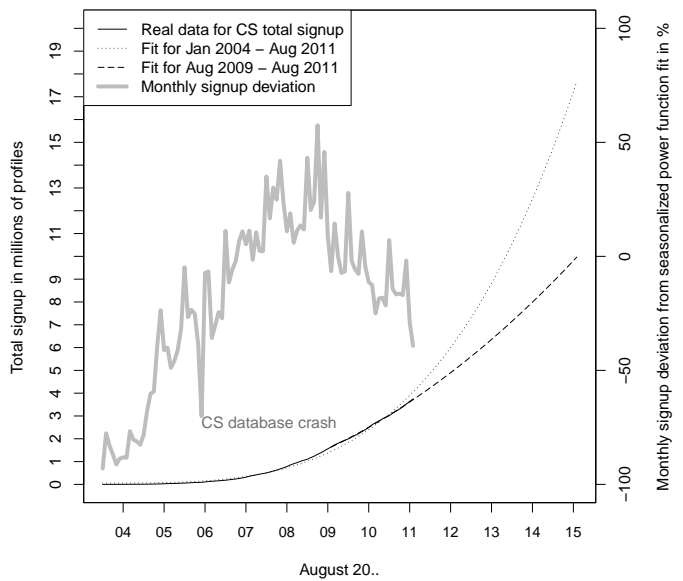}
\caption{Power function fits for CS total signup in Jan 2004 - Aug 2011 and in Aug 2009 - Aug 2011. Monthly CS signup deviation from its seasonalized power function fit as gray curve.}
\label{hccsmigration}
\end{figure}
\indent Progressional Google search volume for the name of a hospex service and the rate of people signing up there are significantly correlated. The correlations coefficients are $.971$ for CS in the years 2005--2011 and $.913$ for BW in the years 2011--2013. Google search volume of BW is about 1-2\% of that of CS. Since we have only two years of Google data for BW and no signup data for HC, we concentrate on CS. Fig.\ref{secondf} shows scaled graphs of monthly Google search and signup for CS. P-value for their linear correlation is below ${10}^{-15}$. First, it allows to make a linear prediction/approximation of now hidden CS signup data based on available Google data. Summing up the approximated monthly signup, we can say with a $99$\% confidence that CS total number of signups was between 6.6M and 6.9M in Dezember 2013, if the linear relationship did not change over the years. Second, we can estimate the HC signup data relying its Google search data or even using Google search data as an equivalent for signup data.\\
\indent Fig.\ref{cskillhc} shows the Google search volume for HC, CS and HC+CS. As you can see, the interest in hospex services is seasonal and growing. Using Box-Cox transformation \cite{boxcox}, we can fit a power function to HC+CS. Averaging the deviations between HC+CS and the fitting power function gives us a growth adjusted average distribution of interest in hospex services over months. August has the highest share with $11.3$\% and December the lowest with $7.1$\%. The five months Jun--Oct have almost the same share as Nov--May. Therefore we use August to mark the time axis on all plots. This seasonality in Google search volume is common for all travel services like Airbnb or Booking.com.\\
\indent You see on fig.\ref{cskillhc} that the interest in CS overtook HC in 2007, whereby the interest in HC started to decrease. Both curves show seasonal variations with peaks in August. The seasonal variations for HC distinctly and visibly flattened from August 2009 on. The growth of the monthly signup for CS decreased from August 2009 on as well. The gray curve on fig.\ref{hccsmigration} shows the deviation between the CS monthly signup and its seasonalized power function fit. On this curve, you see the chasm of the CS database crash in summer 2006 and the clear slowdown after August 2009. This slowdown caused the change in the slope of the total signup curve for CS as depicted on fig.\ref{hccsmigration}. The two power function fits, meaning for Jan 2004 - Aug 2011 and for Aug 2009 - Aug 2011, shape different predictions. The growth of CS users dropped to almost quadratic from being over cubic before August 2009. By the way, the power function fit after August 2009 gives a prediction of 6.9M users for Dezember 2013. Since CS and HC are equivalent services, we assume a user migration period in the years 2006-2009, which powered the faster growth of CS and the decline of HC.\\
\indent The self-proclaimed purpose of the first hospex service Servas is ``world peace, goodwill and understanding by providing opportunities for personal contacts among people of different cultures, backgrounds and nationalities'' \cite{servas}. CS adapted it as ``a better world -- one couch at time'' \cite{cssite}. Combining World Bank and CS data reveals that $75\%$ of the world population, which lives in poor countries with a per capita GDP under \$$10,000$, represented only $10$\% of CS users in 2011. Well, at least for CS the admission to the 'better world' seems to be tightly restricted by income level.\\
\section{Insights for BW}
\begin{figure}
\includegraphics[scale=0.5]{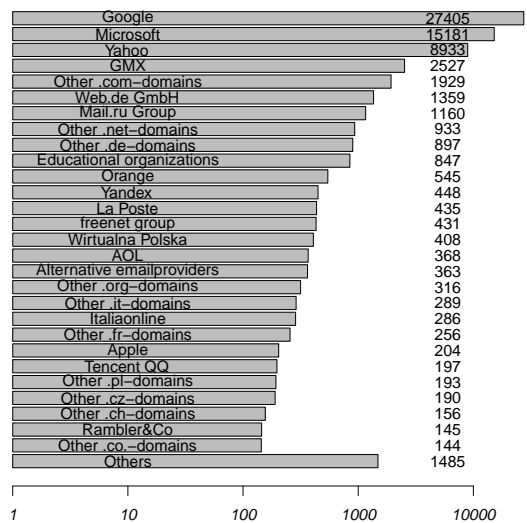}
\caption{Domain name categorization and corresponding histogram in absolute numbers.}
\label{bwemail}
\end{figure}

\begin{figure}
\includegraphics[scale=0.5]{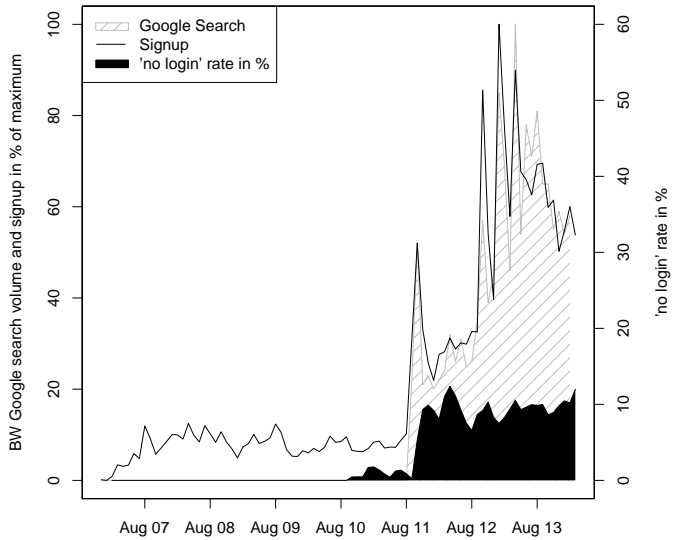}
\caption{BW growth and 'no login' rate.}
\label{nologin}
\end{figure}

\begin{figure}
\includegraphics[scale=0.5]{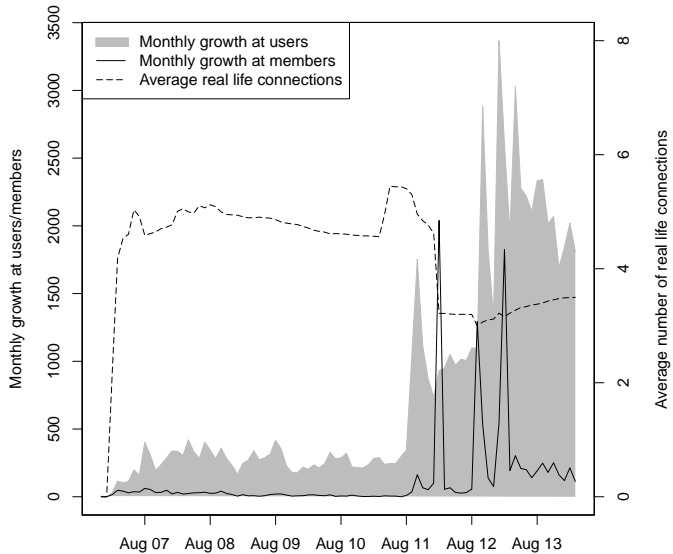}
\caption{Development of BW community.}
\label{bwcom}
\end{figure}

\indent The growth of interest in BW does not display seasonal variation as it is obvious on Fig.\ref{nologin}, but rather is driven by protest of CS users. There is a major peak in BW signups and the google search volume in September 2011 -- CS has been secretly privatized for the personal enrichment of a few a month before \cite{cspriv}. The three peaks between August 2012 and August 2013 are also consequences of changes in CS, which have been considered to be unethical by CS community. The first of these peaks is caused by the update in CS ToU, according to which the CS data can be sold to third parties.\\      
\indent A data set of $68320$ profile entries has been received from BW on 04.03.2014. $68030$ of those entries possess e-mail domain names -- local parts have been removed before data transaction concerning privacy. The remaining $290$ entries are either invalid or officially deleted. Fig.\ref{bwemail} shows the domain name categorization and the corresponding histogram. E-mail services administrated by Google, Microsoft and Yahoo make $75.4$\% of all signups. Domain names of educational organizations like universities and schools from all around the world make $1.2$\%. We have introduced the category of alternative providers, which emphasize privacy and/or certain political goals. These are riseup.[org$|$net], no-log.org, lavabit.com, biomail.de, spamfreemail.de, openmailbox.org, jpberlin.de, mailoo.org, safe-mail.net, immerda.[ch$|$de] and posteo.*. This category makes only $.5$\% of all signups. Although signing up at BW is driven by incorporation of CS and its consequences, majorities choice in email providers is not be explained as ethically dictated.\\
\indent Not every signup at BW leads to subsequent real life interaction. $4921$ out of $68030$ profiles never logged in after signing up. Fig.\ref{nologin} makes a picture of this fact. The rate of users, who did never log in, grew with those signing up after September 2011 and stays at around $10$\%. The incentives for logging in on a hospex website are searching a stay, replying a request and general communication. We consider posting, inviting and contacting members for organization of activities and other minor reasons as examples for general communication. A user interested only in hosting, would not have any incentive to log in until an appropriate hospitality request is forwarded to his e-mail postbox. If communication obviously stimulates login frequency, one is interested in what is stimulating communication.\\
\begin{figure}
\includegraphics[scale=0.5]{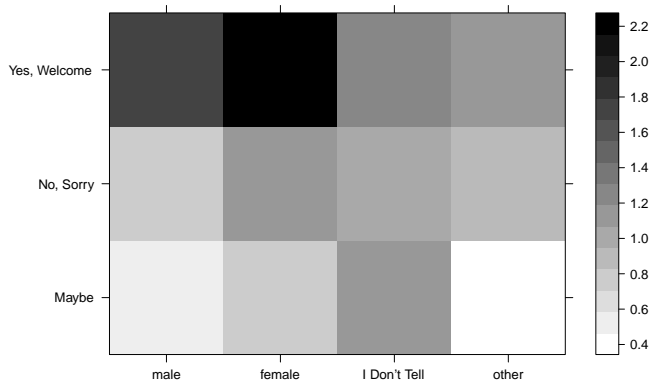}
\caption{Average number of messages received, if neither sent nor posted.}
\label{receiving}
\end{figure}
\indent Under the application of decision tree based feature selection \cite{Weka}, we could determine two factors, which have most impact on number of received messages, in case neither  messages have been sent nor posts have been published. These factors are the indicated gender and the hosting status. Fig.\ref{receiving} shows the average number of messages received depending on these two factors. The profile text was not under investigated features, because it has been replaced by its length to beware users' privacy. For your information, $41.7$\% of BW users indicated to be female and $56.7$\% to be male.\\
\indent Users, who set their hosting status on ``No, sorry'' are considered to be mostly travelers. The reason that they received messages without own initiation are the so-called ``Welcome messages'' sent by BW volunteers. ``Maybe'' is the default case. ``Yes, Welcome'' does not mean that every hospitality request will be accepted -- it only shifts a profile on the top of search results for an entered location.\\
\begin{figure}
\includegraphics[scale=0.5]{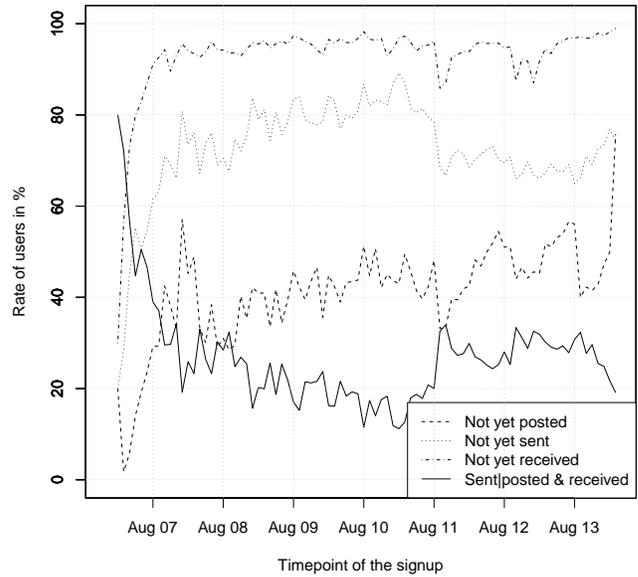}
\caption{Communication rates among BW users.}
\label{comrate}
\end{figure}
\indent Since the text of private messages and public posts has been removed in order to protect users privacy, we can not apply text mining to categorize messages and to determine relationships between them. The correlation between the sum of sent messages and published posts with number of received messages is $.81$ among members. It is higher than the correlation between sent and received messages, which is only $.72$. Therefore, we conclude that publishing posts triggers private messages to some extent. As you can see in fig.\ref{comrate}, the rate of users, which did not receive any message yet, is nearly over $40$\%. This indicates a low share of travelers on BW and/or unequal distribution of existing hospitality requests. The share of users, who use public forums for communication is marginal as you can see in fig.\ref{comrate} -- $98$\% of posts are created by $5$\% of users.\\ 
\begin{table}
\begin{center}
\caption{Distribution of participation in communication.}
\label{distpart}
\begin{tabular}{c|c|c|c}
\hline
                 & neither sent nor posted yet & sent or posted & sum\\
\hline
not yet received &  $43.3$\%                   & $2.7$\%        & $46$\%\\
\hline 
received         &  $26.2$\%                   & $27.8$\%       & $54$\%\\
\hline
sum              &  $69.5$\%                   & $30.5$\%       & $100$\%$\equiv68028$ \\
\end{tabular} 
\end{center} 
\end{table}
\indent Tab.\ref{distpart} shows that the chance to fall into the category of BW users, who sent messages and/or posted, but did not yet receive any private message is low. A calculation on the BW data set of $227695$ messages involving $38425$ users shows that the median of time delay between the first sent/published message/post and first received message is $35$ hours and the average is $68$ days. New users send more than one message, before they received any. This is legitimate, since $68.2$\% of all initiated message exchanges aka conversations are not continued. $36408$ successful conversations between two users at a time resulted a median time delay of $16$ hours and an average of $10$ days. $60$\% of replies arrive in $24$ hours and $86.8$\% of replies arrive in a week. Therefore a time limit of a day means an estimated reply probability of $19.1$\% and of a week $27.6$\%. That means that $10$ private messages to new conversation partners give an estimated probability of $96$\% for receiving at least one reply 
within a week. Archiving the same estimated probability within $24$ hours needs $15$ messages.\\
\indent BW as hospex service does not only facilitate on-line communication, it also creates a hospex community of members interconnected by real life interactions. The only way to estimate the size of this community is to investigate the accommodation reports called comments on BW. The number of real-life interactions is at least as big as the number of comments. A comment can be left by a user for a user and if it has not been disputed, we can assume a real life interaction between those members, even if it is not reciprocated yet. Under this assumption, the size of the BW community was at least $11115$ members by 04.03.2014. Fig.\ref{bwcom} shows the development of the BW community. Monthly growth of the number of users correlates only at $.44$ with the monthly growth of the number of members. The average number of connections per BW member drops after August 2011 and indicates joining of new members into the community.\\ 
\section{Conclusion}
\indent The interest in hospex services is as seasonal as the interest in other accommodation services. Market leadership of a hospex service according to the user number is superable, but takes a relatively long period of time to overcome. The migration period from HC to CS took $4$ years -- from 2006 to 2009. The interest in BW is not seasonal, but correlates with major changes of CS, which are considered to be unethical by the CS community. Most BW users' concerns about ethic of hospex communication facilitating utilities do not apply on choice of e-mail providers. BW lacks at traveler. To archive $96$\% reply probability within a week for an average hospitality request before 04.03.2014, $10$ messages were needed, and $15$ for the same probability within $24$ hours. The average number for real life interactions in BW community variates between $3$ and $5$.\\     
\section*{About and Acknowledgment}
\indent This work was partially published on Bewelcome.org under a Creative Commons License \cite{bwdr}. I want to express my gratitude to the co-members of Bevolunteer for their help. We also thank the people, who provided the Weka library \cite{Weka}.\\
\bibliographystyle{IEEEtran}
\bibliography{bewelcome}

\begin{thebibliography}{10}
\bibitem{servas}
``Servas international,'' servas.org/who-we-are.php.
\end{thebibliography}
\end{document}